\documentclass[twocolumn]{aastex701}

\begin{document}

\title{Site Quality Analysis for an Indian Submillimeter Telescope: A Reanalysis-Based Approach}

\author[orcid=0009-0001-5959-9105, gname=Tanmay, sname=Singh]{Tanmay Singh}
\affiliation{Raman Research Institute, C.~V.~Raman Avenue, Sadashivanagar, Bangalore 560080, India}
\affiliation{School of Earth and Space Exploration, Arizona State University, 781 Terrace Mall, Tempe, AZ 85287, USA}
\email[show]{tsingh65@asu.edu}
\correspondingauthor{Tanmay Singh}

\author[orcid=0000-0002-9761-3676, gname=Mayuri, sname={Sathyanarayana Rao}]{Mayuri Sathyanarayana Rao}
\affiliation{Raman Research Institute, C.~V.~Raman Avenue, Sadashivanagar, Bangalore 560080, India}
\email{mayuris@rri.res.in}

\author[orcid=0000-0002-3351-3078, gname=Ritoban, sname={Basu Thakur}]{Ritoban Basu Thakur}
\affiliation{Department of Physics, California Institute of Technology, Pasadena, CA 91125, USA}
\affiliation{Jet Propulsion Laboratory, Pasadena, CA 91109, USA}
\email{ritoban@caltech.edu}

\begin{abstract}
    
The Himalayan plateau region of Ladakh, India, is a potential host for a science-class submillimeter observatory, building on existing astronomical infrastructure near Hanle and Merak. Using the fifth-generation European Centre for Medium-Range Weather Forecasts (ECMWF) Reanalysis (ERA5) data, we analyze precipitable water vapor (PWV) at monthly resolution over 184 months from January 2010 to April 2025, map PWV statistics across Ladakh, and identify candidate regions that reach PWV $\leq 1$\,mm. For promising locations, we compute atmospheric transmittance and the corresponding atmospheric photon-noise using the \texttt{am} (Atmospheric Model) radiative transfer code; we present transmittance and brightness temperature estimates over $10$--$1000$\,GHz and compare the inferred performance to sites hosting current or planned submillimeter facilities worldwide. We find Ladakh to be favorable for submillimeter observations, with multiple ERA5 grid cells reaching PWV $\leq 1$\,mm. Within ERA5’s spatial resolution, two regions emerge as particularly promising: \texttt{Site A} ($\approx 34.25^{\circ}$N, $78.75^{\circ}$E) and \texttt{Site B} ($\approx 32.50^{\circ}$N, $79.00^{\circ}$E), which satisfy PWV $\leq 1$\,mm for about 23\% and 19\% of the study duration, respectively, compared to about 5\% and 8\% for the Hanle and Merak grid cells. These results motivate targeted \textit{in situ} radiometer measurements for final site selection.

\end{abstract}

\section{Introduction} \label{sec:intro}

Millimetre and submillimeter (mm/submm) wavelengths provide direct access to key astrophysical processes in cold and diffuse environments. 
These include continuum and spectral-line studies of star and planet formation, the structure and chemistry of the cold interstellar medium, and the dust-obscured high-redshift Universe 
\citep{ALMA_partnership,ALMA_disk,Andrews_2020,Benisty_2021}. 
They also enable precision cosmology through measurements of the Sunyaev--Zeldovich (SZ) effect 
\citep{SunyaevZeldovich1970,Birkinshaw1999,Carlstrom2002}, and provide unique diagnostics of solar and stellar atmospheres 
\citep[e.g.][]{White2017ALMA_sun}.

Ground-based submillimeter astronomy is, however, strongly limited by atmospheric absorption, primarily due to water vapor 
\citep{li_2017}. 
The vertically integrated column of water vapor is quantified as precipitable water vapor (PWV), which governs both atmospheric transmission and sky brightness temperature. 
While science requirements vary, PWV values of $\lesssim 1$\,mm are generally considered favorable, with lower values enabling sustained access to higher-frequency submillimeter windows.

Submillimeter facilities are therefore located at the highest and driest sites on Earth. 
This is evidenced by the PWV at existing submillimeter observatories. 
The Llano de Chajnantor plateau in Chile (ALMA/AOS, $\sim$5050\,m) typically exhibits median PWV of $\sim$1.0--1.1\,mm, with PWV $<0.5$\,mm up to $\sim$25\% of the time; the nearby Cerro Chajnantor summit (5612\,m) is even drier, showing $\approx$28\% lower PWV in long-term statistics and $\sim$36\% lower in simultaneous campaigns 
\citep{ALMA_partnership,Cortes_2020,Bustos2014Parque,Radford2016PASP,Pardo_2022,Pardo_2025}. 
The Antarctic plateau (South Pole, Dome C, Dome A) provides exceptionally low and stable PWV 
\citep{Chamberlin2001JGR,Lawrence2004,Sims2012DomeA}. 

Beyond low PWV, effective site selection must also account for the stability of atmospheric transmission and sky brightness temperature. 
Rapid water-vapor-induced fluctuations generate excess sky noise, which can dominate both bolometric and heterodyne system noise budgets; for example, the South Pole shows significantly reduced sky noise relative to mid-latitude sites under comparable transparency 
\citep{LayHalverson2000}. 
Increased altitude further mitigates pressure broadening and reduces dry-air continuum contributions, improving both mean transmission and temporal stability across submillimeter frequencies 
\citep{Pardo2001ATM,Paine2022am,Matsushita2017PASP}. 

Among northern-hemisphere high-altitude sites, Mauna Kea remains one of the best-characterized benchmarks for submillimeter transmittance. 
Long-term tipping radiometer measurements indicate median 225\,GHz zenith optical depths corresponding to PWV values of order 1--2\,mm under favorable conditions 
\citep{Mason1994,Naylor2000MaunaKea,Radford2016PASP}. 
These measurements provide a valuable northern reference for atmospheric transparency comparisons. 
At the same time, long-term campaigns in the southern hemisphere, particularly at Chajnantor, have used radiometers and heterodyne instruments to quantify atmospheric transmission, seasonal variability, and access to the driest submillimeter windows 
\citep[e.g.][]{Cortes_2020,Radford2016PASP}.

Within this global context, the availability of a dry northern-hemisphere site would complement southern facilities by extending high-frequency submillimeter coverage to northern targets.

The high-altitude Himalayan plateau extends across northern India and into the trans-Himalayan region of Ladakh. 
With an average elevation of $\sim3000$\,m above mean sea level \citep{Ningombam_2015}, the combination of altitude, arid climate, and reduced anthropogenic activity provides observing conditions propitious to high-quality multi-wavelength astronomy. 
Existing facilities in this region demonstrate its scientific viability. 
The Indian Astronomical Observatory \citep{cowsik2002introduction} at Hanle (4500\,m, 32.7789$^{\circ}$\,N, 78.9651$^{\circ}$\,E) hosts several telescopes, including the Himalayan Chandra Telescope (HCT), the HAGAR array \citep{chitnis2011status}, and the MACE gamma-ray telescope \citep{MACE}. 
The National Large Solar Telescope is located at Merak (33.79$^{\circ}$\,N, 78.61$^{\circ}$\,E), while in China the Ali CMB Polarization Telescope (AliCPT) is situated on the Tibetan Plateau at an elevation of $\sim5250$\,m \citep{AliCPT}. 

Given its cold desert environment, high elevation, and extended clear winter season outside the monsoon period 
\citep{Ningombam2020Hanle}, Ladakh has been considered a promising candidate for hosting a northern-hemisphere submillimeter facility 
\citep{Ananthasubramanian2001Hanle,TKS}.

In this work, we assess the suitability of multiple candidate sites in the Ladakh region for a future Indian submillimeter observatory. 
We analyse the spatio-temporal variability of PWV across representative pixels and examine its impact on atmospheric transmittance and zenith sky brightness in order to identify regions that provide the most favourable conditions for high-frequency submillimeter observations.

In Section~\ref{sec:methods} we describe the datasets and PWV estimation procedures. 
Section~\ref{sec:main_results} presents the results of this study followed by radiative transfer modeling and resulting atmospheric transmission properties. 
Section~\ref{section:discussion} interprets the results in terms of site performance and feasibility for a submillimeter observatory.

\section{Methods}\label{sec:methods}

In this study, we conduct a preliminary exploration over a relatively large geographical region to identify potential sites for submillimeter facilities. As detailed in situ opacity measurements with dedicated and cross-calibrated instruments are impractical across the entire Ladakh plateau, we first use available large-scale datasets, for example atmospheric reanalysis products, to screen the region and identify a small number of promising candidate areas. These shortlisted locations can then be evaluated in a second stage through targeted on-site measurements.

\subsection{Reanalysis Dataset}\label{subsec:data}

We use global atmospheric reanalysis products to identify candidate sites on the Ladakh plateau that are suitable for a science-quality submillimeter telescope. Reanalysis datasets blend heterogeneous observations with a global numerical model via data assimilation to produce spatially and temporally complete fields; this reduces random errors and sampling gaps but does not remove all model or observational biases \citep{Hersbach2020ERA5,Hersbach_ERA5_monthly_pressure_2023}. 

ERA5 is the fifth generation ECMWF (European Centre for Medium-Range Weather Forecasts) reanalysis data set produced using the CY41R2 version of the Integrated Forecast System (IFS). It employs a four-dimensional variational data assimilation scheme (4D-Var), in which observations distributed in space and time are assimilated over a finite time window by optimizing the atmospheric state to minimize the mismatch between model forecasts and observations, subject to the governing physical equations \citep{Hersbach2020ERA5,lavers2022evaluation}. This approach allows ERA5 to account not only for spatial consistency but also for temporal evolution, making it particularly suitable for studies of atmospheric variability relevant to site characterization.

ERA5 incorporates data from modern multi-satellite sounders and imagers as well as conventional ground-based observations and has demonstrated robust agreement with independent measurements across a range of climatic regimes \citep{Huang_2021}. It provides atmospheric fields at 37 pressure levels from 1000\,hPa to 1\,hPa, with products available at both hourly and monthly temporal resolution. ERA5 also offers among the highest spatial resolution of current global reanalysis datasets, at $0.25^\circ \times 0.25^\circ$ (longitude $\times$ latitude), corresponding to approximately 23\,km $\times$ 27\,km over the Ladakh region. This relatively fine resolution is particularly important for resolving the complex topography of the plateau and for distinguishing between nearby potential sites. We use the monthly averaged ERA5 products, as our primary objective is to capture long-term climatological trends relevant to preliminary site selection.

\subsection{Water Vapor Measurements}
A 220\,GHz tipping radiometer has operated at the Indian Astronomical Observatory (IAO), Hanle (Ladakh), since late 1999, with published datasets spanning 2000–2003 and 2006–2018 \citep{Ananthasubramanian2001Hanle,Ningombam2020Hanle}. The radiometer provides in situ measurements of atmospheric transparency that can be converted to precipitable water vapor (PWV) and thus serve as a reference for validating reanalysis-based PWV at Hanle (e.g. ERA5; MERRA-2) \citep{Hersbach2020ERA5,Gelaro2017MERRA2}.

The instrument determines atmospheric opacity by observing the sky brightness temperature at multiple elevation angles (``sky dips'') and fitting a plane-parallel transmission model to retrieve the zenith optical depth, which is then related to PWV \citep{Chamberlin1994AO,Chamberlin1995IJIMW,ThomasOsip2007PASP}. The dataset analysed by \citet{Ningombam2020Hanle} reports monthly median PWV (via 220\,GHz opacity) from September~2006 to October~2018, capturing the seasonal cycle and interannual variability. We use these measurements to validate the reanalysis-based PWV estimations.

\subsection{PWV Estimation from ERA5 reanalysis data}\label{subsec:pwv_calc}

\begin{figure}
    \resizebox{\hsize}{!}{%
        \includegraphics[scale=0.7]{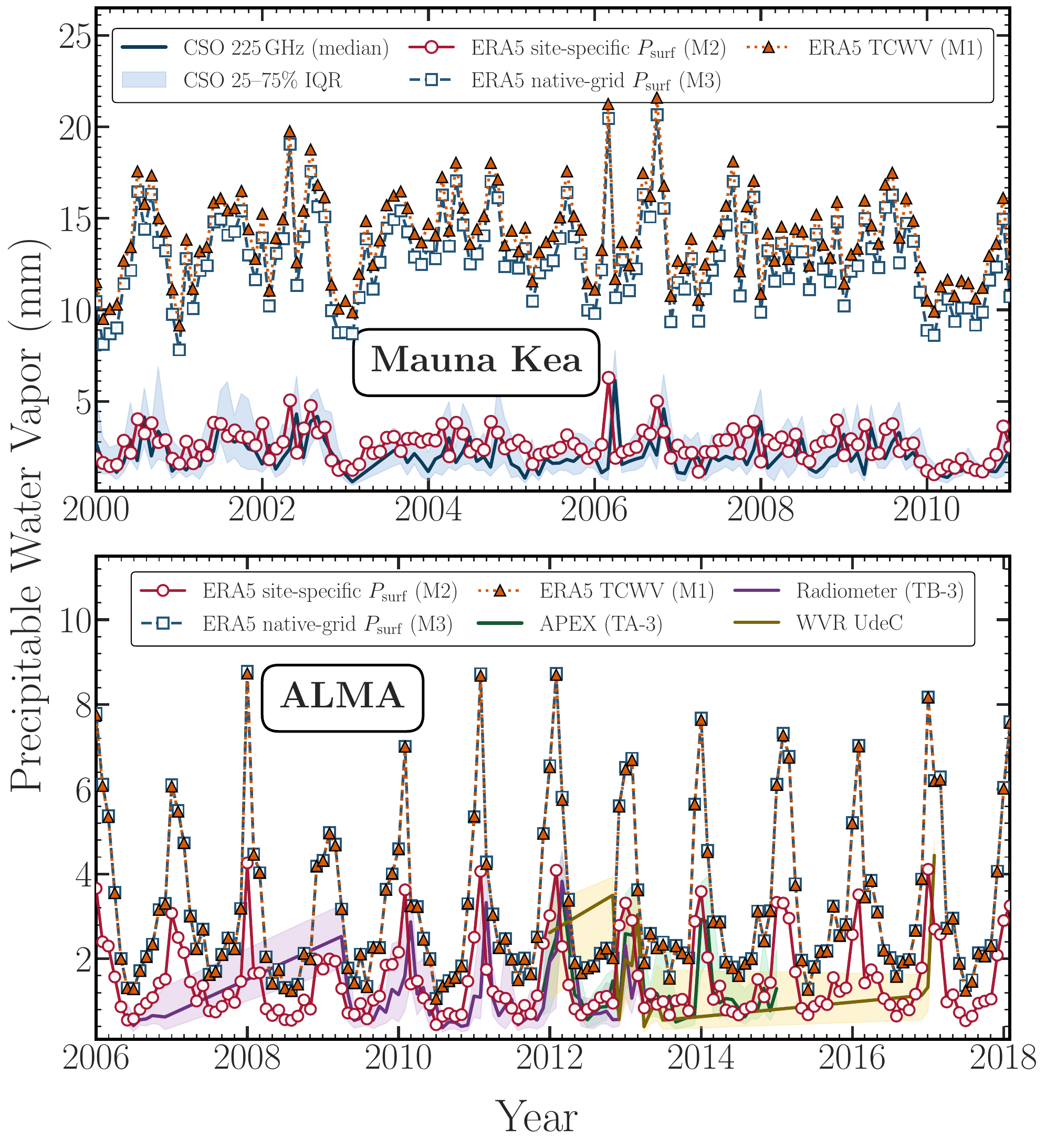}%
    }
    \caption{
        Comparison of precipitable water vapor (PWV) values derived from different methods with radiometer measurements at Mauna Kea (top) and the ALMA site at Chajnantor (bottom). 
        Red triangles, magenta circles and blue squares correspond to the PWV values derived from the Methods $M1$, $M2$ and $M3$ respectively as described in this work. All other markers denote radiometer measurements compiled from \citet{Valeria_2024} for Mauna Kea and \citet{Cortes_2020} for the ALMA site.
    }
    \label{fig:mauna_kea_alma_comparison}
\end{figure}

The troposphere, extending from Earth's surface up to $\simeq 10$ km altitude, contains around 99 percent of atmospheric water vapor \citep[e.g.][]{Shikhovtsev_2022,WallaceHobbs2006}. While ERA5 provides precipitable water vapor (PWV) directly as Total Column Water Vapor (TCWV) \citep{C3S_ERA5_PLMM,C3S_ERA5_Single,Hersbach2020ERA5}, PWV can also be analytically computed from specific humidity \( q \) and atmospheric pressure \( p \) as:

\begin{equation}\label{eq:PWV}
\text{PWV}(\lambda, \phi, t) = \frac{1}{\rho_w g} \int_{p_{\text{top}}}^{p_s(\lambda, \phi,t)} q(\lambda, \phi, p,t)\; dp,
\end{equation}

where \( \lambda \), \( \phi \), and \( t \) represent longitude, latitude, and time, respectively. Here, \(\rho_w\) is the density of liquid water (1000 kg\,m$^{-3}$), \( g \) is gravitational acceleration (9.81 m\,s$^{-2}$), \( p_s(\lambda,\phi,t) \) is the surface pressure at the given site, and \( p_{\text{top}} \) is the upper integration limit, set to 1\,hPa, above which water vapor contribution is considered negligible.

ERA5 provides \( q(\lambda,\phi,p) \) as monthly mean values at discrete pressure levels and also provides integrated PWV as TCWV. Because ERA5 uses a finite-resolution grid, each value is the area average of a model cell and is stored at the cell center (a “pixel”). When estimating PWV at a specific site, especially in steep terrain, using the ERA5 grid-cell surface pressure (\texttt{sp}) as the lower limit in Eq.~\ref{eq:PWV} can introduce large errors \citep[e.g.,][]{Huang_2021,Wei2025_Mongolia_ERA5}. A good example is the Mauna Kea summit (elevation $\sim$4200\,m), where directly using the nearest or interpolated ERA5 grid-point surface pressure in Equation~\ref{eq:PWV} produces unrealistically high PWV values ($\sim 10$\,mm), inconsistent with in situ measurements \citep{vanKooten_2022}. This discrepancy arises because the lowest atmospheric layers contain the bulk of water vapor, which likely varies strongly within the area associated with the grid point. Thus, the choice of the lower integration limit \( p_s \) critically affects the estimated PWV, and even small deviations from the actual site surface pressure lead to substantial errors. We therefore list three methods for PWV estimation in ERA5 or similar reanalysis datasets: (i) using ERA5's native Total Column Water Vapor (TCWV) (ii) a site-specific approach for cases where the exact coordinates and elevation or surface pressure profile are known, and (iii) a pixel-averaged approach for analyses conducted at an averaged and coarser spatial scale.

\begin{figure}[t]
    \resizebox{\hsize}{!}{\includegraphics[trim={0cm, 0cm, 0cm, 0cm}, clip, width=15cm]{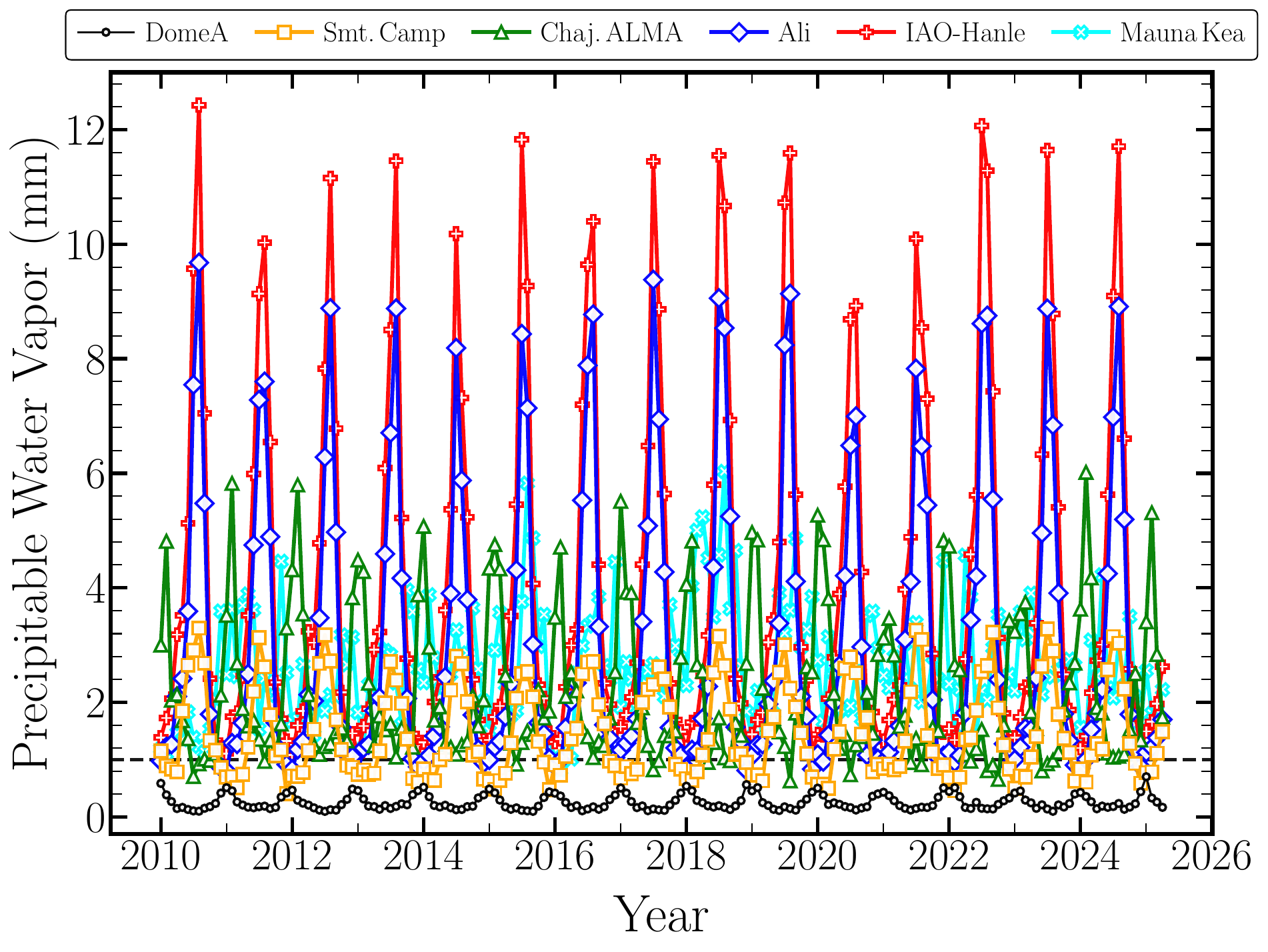}}
    \caption{The monthly average PWV across global sites over 2010 to April 2025 calculated using $M2$. Dome A has the best PWV, consistently remaining below 1 mm across all months over the full timescale.}
    \label{fig:global_sites}
\end{figure}

\paragraph{Method~1 $(M1)$: TCWV data product } 
In this method, PWV is directly obtained from the ERA5 Total Column Water Vapor (TCWV) variable. TCWV represents the vertically integrated water vapor column for each ERA5 grid cell, computed internally by the reanalysis system using its native surface pressure and humidity structure.

\paragraph{Method~2 $(M2)$: Site-specific (geopotential-height interpolation).}
To obtain accurate PWV for known observatory sites with precise \((\lambda,\phi,H_{\text{site}})\), we determine the site-specific surface pressure \(p_s\) from ERA5 geopotential height-pressure-level relationship at the site elevation and then integrate the ERA5 specific-humidity profile down to \(p_s\) using Eq.~\ref{eq:PWV}.
This approach relies directly on ERA5 monthly pressure level fields and proceeds as follows:
\begin{itemize}
\item \textit{Horizontal (spatial) interpolation:}
    At each time step we extract the ERA5 monthly mean pressure–level fields. These are the pressure values \(p_k\), the geopotential heights \(Z_k\) and the specific humidity \(q_k\). The index \(k = 1,\dots,N_p\) labels the discrete pressure levels in the model. For a site at longitude \(\lambda\) and latitude \(\phi\) we perform bilinear interpolation in longitude and latitude at each pressure level using the four surrounding ERA5 grid points. This gives a site–centred vertical column on the native ERA5 pressure grid with geopotential height \(Z_k(\lambda,\phi)\) and specific humidity \(q_k(\lambda,\phi)\) for each month in the dataset.
    \item \textit{Elevation-based interpolation:}
    For each interpolated monthly profile, we identify the two consecutive ERA5 pressure levels, $p_k$ and $p_{k+1}$, in the vertical column whose geopotential heights $Z_k$ and $Z_{k+1}$ bracket the site elevation $H_{\text{site}}$ ($Z_k \le H_{\text{site}} \le Z_{k+1}$). We then estimate the site-specific surface pressure $p_s$ by logarithmic interpolation, assuming an exponential pressure variation within this layer,
    \begin{equation}
    \ln p_s
    = \ln p_k
    + \frac{H_{\text{site}} - Z_k}{Z_{k+1} - Z_k}
    \left(\ln p_{k+1} - \ln p_k\right),
    \end{equation}
    which follows from the hypsometric relation under a layer-mean virtual-temperature assumption \citep{WallaceHobbs2006,HoltonHakim2013}.
    
\item \textit{Vertical interpolation:}
Since the computed surface pressure $p_s$ does not generally coincide with a native ERA5 pressure level, we perform a monotonic, shape-preserving vertical interpolation of the specific humidity profile $q(p)$ to the exact pressure $p_s$ using the Piecewise Cubic Hermite Interpolating Polynomial (PCHIP) \citep{FritschCarlson1980,FritschButland1984}. The pressure-integrated water vapor column is then evaluated via the following sub-steps:
\begin{itemize}
    \item The layer-averaged specific humidity and corresponding pressure increments are defined as
    \begin{equation}
    \bar{q}_i = \frac{q_i + q_{i+1}}{2},
    \qquad
    \Delta p_i = p_i - p_{i+1}.
    \end{equation}

    \item The vertically integrated water-vapor column is then obtained by integrating from the model top pressure $p_{\text{top}}$ down to the site-specific pressure $p_s$. The contribution from the resolved ERA5 levels and the residual layer between the highest level $p_m \leq p_s$ and $p_s$ are combined to yield
    \begin{equation}
    \mathrm{PWV}
    = \frac{1}{\rho_w g}
    \left[
    \sum_{i=1}^{m-1} \bar{q}_i \, \Delta p_i
    + q_{\text{interp}} (p_s - p_m)
    \right].
    \end{equation}
\end{itemize}

Representativeness errors arise when a coarse ERA5 grid cell, defined by its mean elevation and grid-averaged humidity profile, is used to describe the atmospheric column above a specific observatory located on a ridge or plateau. By adjusting the column to the true site pressure \(p_s\) and integrating humidity along this corrected vertical coordinate, the resulting PWV more accurately reflects the column that an instrument at the site would sample, compared to the unadjusted grid-cell PWV.
\end{itemize}

\paragraph{Method~3 $(M3)$: Regional pixel survey (native-grid approach).}
For a comprehensive, region-wide assessment (e.g., across the Ladakh plateau), we compute PWV at the native ERA5 grid. In this approach, we use the ERA5 surface pressure (\texttt{sp}) provided for each grid cell directly as the lower integration bound in Eq.~\ref{eq:PWV}, and integrate the ERA5 monthly-mean \(q(p)\) profiles accordingly at each grid point. This pixel-based method characterizes average conditions representative of the grid cell (rather than any specific micro-site), which is appropriate for first-pass regional screening and ranking of candidate areas prior to targeted on-site campaigns. This approach is conceptually similar to $M1$ with interpolation implemented for improved accuracy. Regional analysis of pixel-level PWV values and duty-cycle statistics for the Ladakh region is presented in Section ~\ref{sec:results}.

To illustrate the practical consequences of the three PWV estimation approaches introduced above, we compare them with independent radiometer measurements at two benchmark observatories (Mauna Kea and Chajnantor/ALMA) in Fig.~\ref{fig:mauna_kea_alma_comparison}. Each panel shows three ERA5-based series alongside the observations: ERA5 TCWV ($M1$, red triangles), the site-specific integration using geopotential-derived \(p_s\) ($M2$, magenta circles), and the native-grid integration using grid-cell surface pressure ($M3$, blue squares). The remaining curves and markers represent radiometer data products (\citealt{Valeria_2024}; \citealt{Cortes_2020}). Two key features emerge. First, the TCWV values and the native-grid integration ($M1$ and $M3$) are systematically high at elevated sites, since they implicitly include excess low-level moisture absent from the true atmospheric column above the observatory. Second, enforcing the site altitude through geopotential interpolation ($M2$) yields substantially better agreement with the radiometers, particularly at Mauna Kea where orographic representativeness errors are most severe \citep{Otarola2010,Cortes_2020,RadfordChamberlin2000}. The TCWV method ($M1$) is not used further in this study. Since $M2$ provides the most reliable results, it is adopted for all subsequent site-specific  (exact latitude-longitude) based analyses. $M3$ is used for pixel-based, coarse regional study. 

Using $M2$, we place our methodology in a broader context in Fig.~\ref{fig:global_sites} which shows monthly PWV time series (2010–April~2025) for a set of widely used high–altitude observatory locations. All sites exhibit a seasonal cycle, with winter minima and summer maxima. As expected, Dome~A remains consistently below the \(\sim\)1\,mm threshold throughout the record, whereas Summit Camp, Mauna Kea, and Chajnantor/ALMA span \(\sim\)0.5–6\,mm depending on season. Relative to the other global comparison sites, IAO--Hanle shows systematically higher PWV values, though it remains most comparable to the Ali site, another high-altitude site on the Tibetan Plateau, and reaches a minimum monthly PWV of 1.05 mm during the study period. This global comparison accomplishes two goals: it (i) establishes the \(\mathrm{PWV}\le1\)~mm benchmark that we adopt for screening duty cycle and atmospheric windows relevant to sub–mm astronomy, and (ii) motivates our strategy of analyzing both \emph{full–year} and \emph{winter–only} conditions in the Ladakh region. In what follows, we therefore use the validated site-specific method for point sites and the native-grid method for regional analysis.

\setlength{\tabcolsep}{2pt}  

\centerwidetable

\begin{deluxetable*}{lccccccccccccccccc}
\tabletypesize{\scriptsize}          
\tablewidth{\textwidth}            

\tablecaption{Geographic coordinates and monthly precipitable
water-vapor (PWV) statistics (2010–2025) for candidate submillimeter
astronomy sites.\label{table:pwv_sites}}

\tablehead{
\colhead{Site} &
\colhead{Lat\textsuperscript{(1)}} &
\colhead{Lon\textsuperscript{(2)}} &
\colhead{Elev\textsuperscript{(3)}} &
\colhead{$N$\textsuperscript{(4)}} &
\colhead{$\mu$\textsuperscript{(5)}} &
\colhead{$\tilde{x}$\textsuperscript{(6)}} &
\colhead{SD\textsuperscript{(7)}} &
\colhead{Min\textsuperscript{(8)}} &
\colhead{Max\textsuperscript{(9)}} &
\colhead{$P_{05}$\textsuperscript{(10)}} &
\colhead{$P_{25}$\textsuperscript{(11)}} &
\colhead{$P_{75}$\textsuperscript{(12)}} &
\colhead{$P_{95}$\textsuperscript{(13)}} &
\colhead{CV\textsuperscript{(14)}} &
\colhead{$f_{<0.5}$\textsuperscript{(15)}} &
\colhead{$f_{<1}$\textsuperscript{(16)}} &
\colhead{$f_{<2}$\textsuperscript{(17)}}}

\startdata
Ali         &  32.63 &  80.00 & 5100 & 184 & 2.99 & 1.75 & 2.51 & 0.81 &  9.68 & 0.89 & 1.17 & 4.22 &  8.73 & 0.84 & 0.00 & 0.13 & 0.55 \\
Chaj.\ ALMA & -23.02 & 292.25 & 4800 & 184 & 2.19 & 1.66 & 1.29 & 0.63 &  6.02 & 0.92 & 1.27 & 2.71 &  4.83 & 0.59 & 0.00 & 0.11 & 0.60 \\
Dome\,A     & -80.37 &  77.35 & 4091 & 184 & 0.26 & 0.21 & 0.13 & 0.10 &  0.70 & 0.12 & 0.16 & 0.35 &  0.51 & 0.50 & 0.94 & 1.00 & 1.00 \\
IAO--Hanle  &  32.78 &  78.96 & 4500 & 184 & 4.01 & 2.48 & 3.17 & 1.08 & 12.43 & 1.26 & 1.69 & 5.63 & 11.26 & 0.79 & 0.00 & 0.00 & 0.38 \\
Mauna Kea   &  19.82 & 204.53 & 4200 & 184 & 2.66 & 2.51 & 0.98 & 0.82 &  6.03 & 1.26 & 1.95 & 3.27 &  4.51 & 0.37 & 0.00 & 0.01 & 0.26 \\
Summit Camp &  72.57 & 321.54 & 3694 & 184 & 1.48 & 1.17 & 0.81 & 0.41 &  3.30 & 0.59 & 0.79 & 2.12 &  3.03 & 0.55 & 0.03 & 0.40 & 0.73 \\
\enddata

\tablecomments{(1) Lat (\textdegree N) (2) Lon (\textdegree E) (3) Elevation (m)
(4) Monthly samples, January 2010–April 2025 (5) Mean PWV (mm)
(6) Median PWV (mm) (7) Standard deviation in PWV (mm);
(8) Minimum PWV (mm) (9) Maximum PWV (mm) (10–13) 5th, 25th,
75th, 95th percentiles (14) Coefficient of variation (15–17) Fraction
of months with PWV $\le 0.5$, $\le 1$, $\le 2$ mm.}
\end{deluxetable*}

\setlength{\tabcolsep}{6pt}

\section{Results}\label{sec:main_results}

\subsection{ERA5 Validation at IAO Hanle}\label{subsec:hanle_validation}

\begin{figure}
    \resizebox{\hsize}{!}{\includegraphics[scale=0.5]{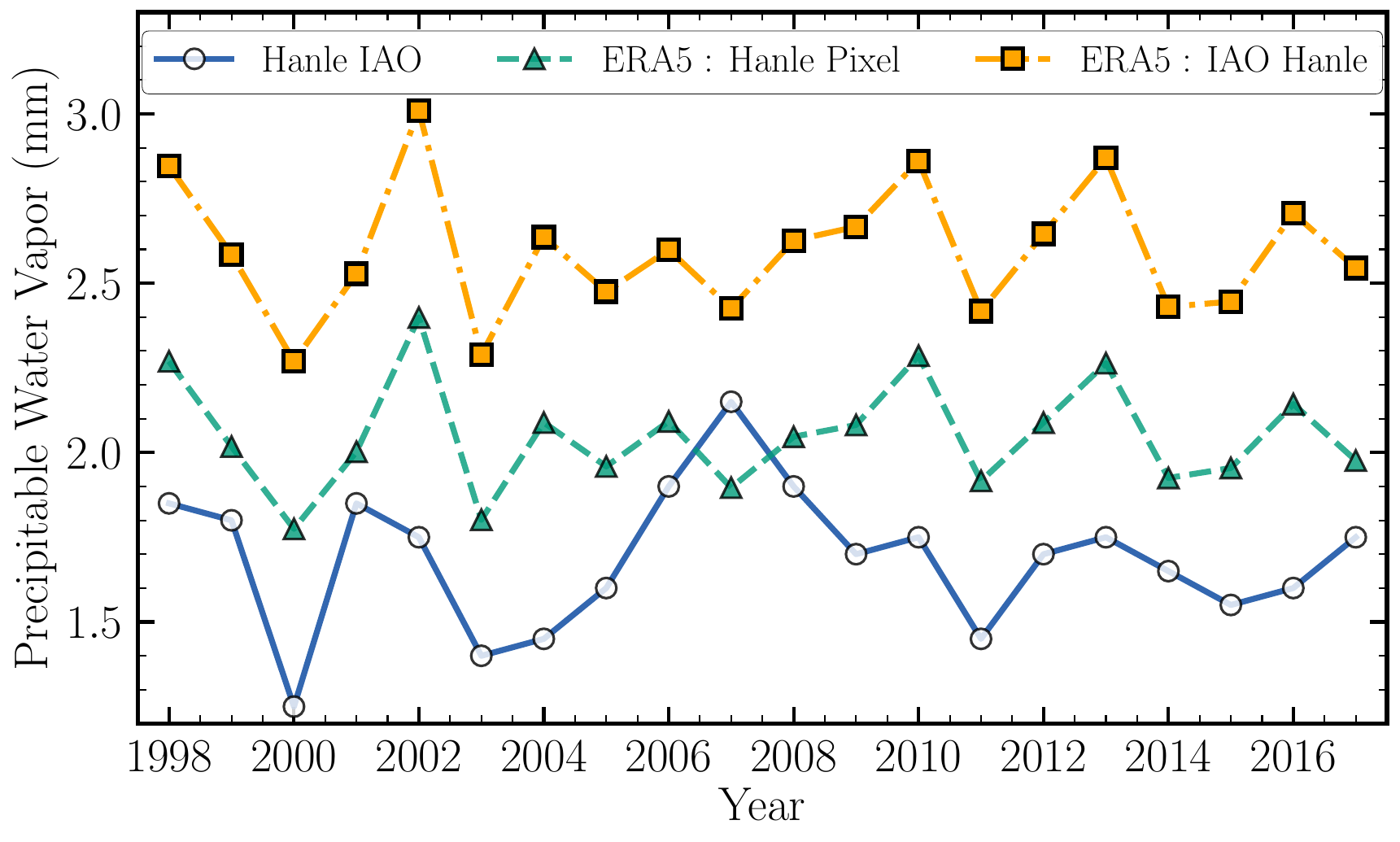}}
    \caption{Annual median PWV at the Indian Astronomical Observatory (IAO, Hanle) for 1998--2017. The plot compares radiometer measurements (IAO--Hanle; blue-white markers) with ERA5-derived PWV computed using the site-specific method based on geopotential-height–interpolated surface pressure (\emph{ERA5: IAO--Hanle}, M2; orange-black markers) and the native-grid method using the nearest ERA5 grid cell (\emph{ERA5: Hanle Pixel}, M3; green markers). ERA5 systematically overestimates PWV relative to the radiometer, but reproduces the observed interannual variability and long-term trend.}
    \label{fig:Hanle_era_obs_yearly}
\end{figure}

\begin{figure}[htbp]
    \resizebox{\hsize}{!}{\includegraphics[scale=0.5]{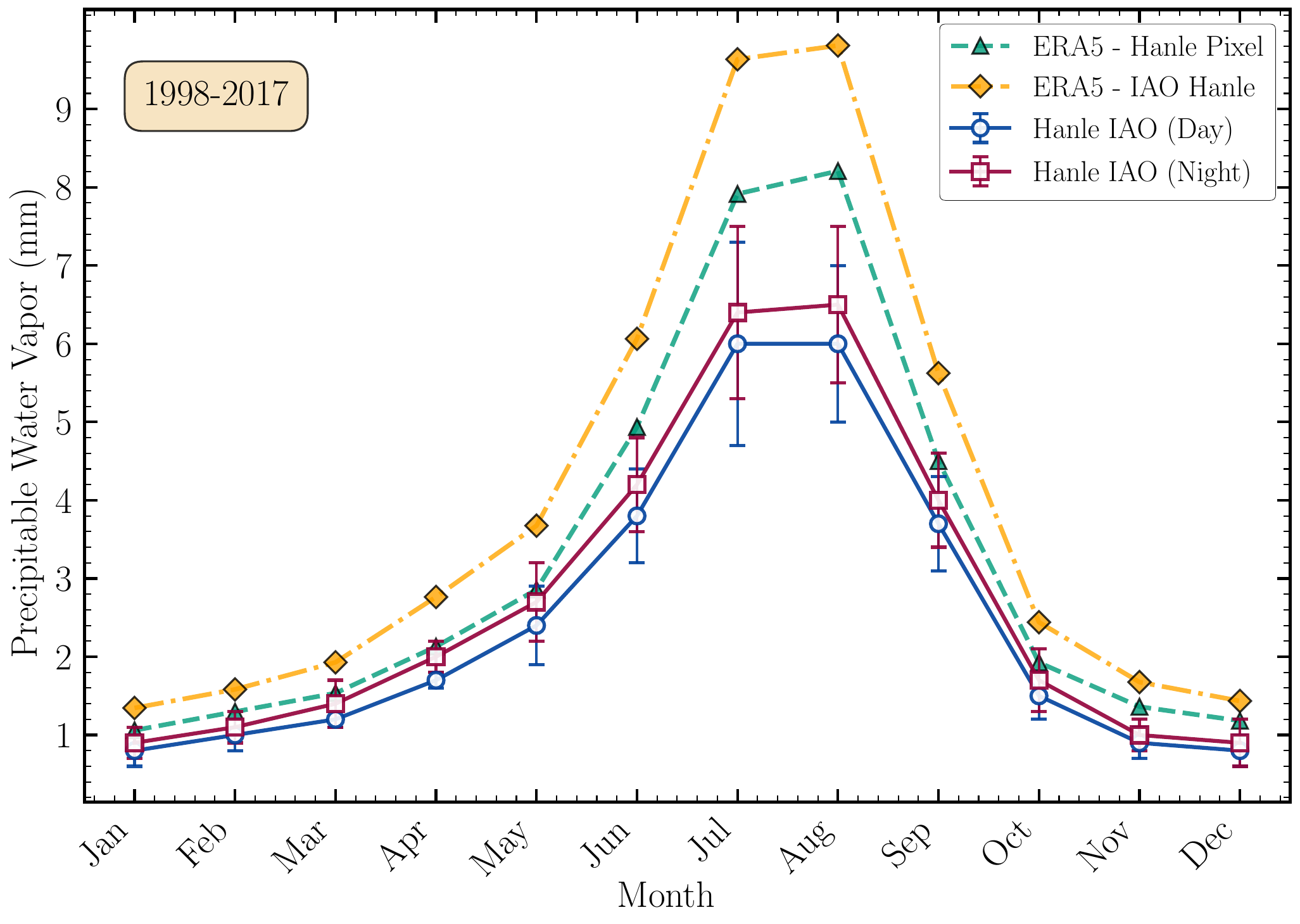}}
    \caption{Monthly median PWV climatology at IAO--Hanle for 1998--2017. Daytime (blue) and nighttime (red) radiometer medians are compared with the ERA5 site-specific series (\emph{ERA5: IAO--Hanle}, M2; orange symbols) and, for reference, the native-grid series (\emph{ERA5: Hanle Pixel}, M3; green symbols). The driest conditions (PWV $\sim 1$\,mm or lower) occur during winter (November--February), during which the ERA5 site-specific series generally lies within the observational range, while larger positive departures are evident in the wetter months (April--October).}
    \label{fig:Hanle_era_obs_monthly}
\end{figure}

To assess the fidelity of the PWV calculation procedures outlined in Section~\ref{subsec:pwv_calc}, we compare ERA5-derived PWV at the exact IAO--Hanle site calculated using the site-specific method (M2) against on-site tipping radiometer measurements for 1998--2017. Figure~\ref{fig:Hanle_era_obs_yearly} shows annual medians for the radiometer series (\emph{IAO--Hanle}) alongside two ERA5 realizations: the \emph{site-specific} series (\emph{ERA5: IAO--Hanle}, M2) and, for comparison, the \emph{native-grid} series (\emph{ERA5: Hanle Pixel}, M3). The upper panel indicates that ERA5 systematically exceeds the radiometer medians, a pattern made explicit in the residuals shown in the lower panel. Despite this positive offset, the \emph{ERA5: IAO--Hanle} series reproduces the interannual variability and overall trend evident in the observations, supporting its use as a conservative (upper-bound) estimate of atmospheric moisture for site assessment.

Figure~\ref{fig:Hanle_era_obs_monthly} presents the monthly median PWV climatology, with radiometer data separated into day and night and compared with the ERA5-derived monthly medians from the \emph{site-specific} series (\emph{ERA5: IAO--Hanle}, M2) and, for reference, the \emph{native-grid} series (\emph{ERA5: Hanle Pixel}, M3). During the dry winter months (November--February), observed medians are typically close to 1\,mm and the ERA5 site-specific series lies mostly within the observational uncertainties. In contrast, during the wetter months (April--October), ERA5 exhibits a larger positive offset, consistent with known difficulties of reanalysis in relatively humid conditions over complex orography on the Tibetan Plateau/High Mountain Asia 
\citep[see e.g.][]{Huang_2021,He2021ESS,Ren2021AtmosQTP,Ou2023NWTP}.

Quantitatively, relative to the radiometer, the \emph{ERA5: IAO--Hanle} annual series exhibits a systematic positive bias of 0.904\,mm, an RMSE of 0.929\,mm, and a modest interannual Pearson correlation of $r=0.397$. The radiometer indicates a weakly negative PWV trend of $-0.006$\,mm\,decade$^{-1}$, whereas \emph{ERA5: IAO--Hanle} is nearly flat at $+0.001$\,mm\,decade$^{-1}$. On the monthly climatology (computed on the mean of day and night), \emph{ERA5: IAO--Hanle} yields a bias of 0.765\,mm, an RMSE of 0.859\,mm, and a very high correlation ($r=0.996$). A first-harmonic fit to the seasonal cycle shows negligible phase error (phase shift $=-0.04$\,months), but a pronounced amplitude overestimation ($\sim 52\%$ larger than observed), in line with the summer-season residuals.

For completeness, the interpolated gridcell series (\emph{ERA5: Hanle Pixel}) achieves smaller error scores against the radiometer (monthly RMSE $\approx 0.372$\,mm; yearly RMSE $\approx 0.345$\,mm). However, this apparent improvement is a representativeness artifact as the gridcell surface pressure at the Hanle pixel (approximately 555\,hPa) is lower than the true site-specific value (approximately 585\,hPa) derived from geopotential-height interpolation and also confirmed by observational studies \citet{Ningombam_2020b}, thereby shortening the vertical integration column and suppressing ERA5 PWV. The reduced errors thus reflect a tradeoff between humidity bias and an erroneously shallow column, rather than reanalysis model values. In summary, despite the positive bias, the \emph{ERA5: IAO--Hanle} ($M2$) series preserves the correct temporal variability and seasonal phase, confirming its suitability as a conservative upper-bound estimate for site characterization as discussed in Section~\ref{subsec:pwv_calc}.

\subsection{Exploring Ladakh}\label{sec:results}

\begin{figure}
\includegraphics[trim={0cm, 0cm, 0cm, 0cm}, clip, width=9cm]{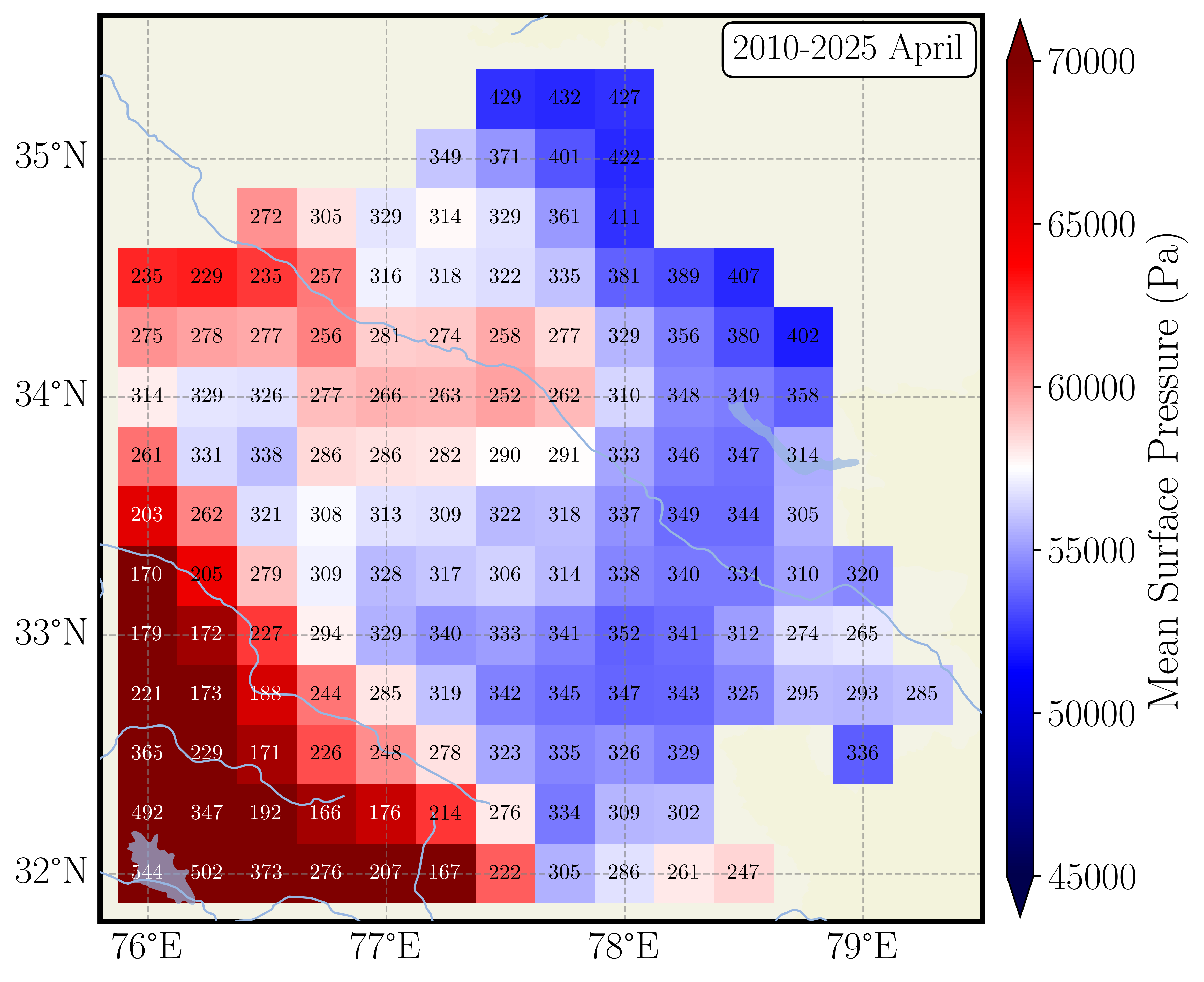}
\caption{
ERA5 gridded mean surface pressure map averaged over the period from 2010 to April 2025, used for computing precipitable water vapor (PWV) at each grid point within the region of interest. The standard deviation of surface pressure at each grid is noted in each pixel.
}
\label{fig:pressure_map}
\end{figure}

\begin{figure}
\centering
\includegraphics[trim={0.0cm 0cm 0cm 0cm},clip,width=9cm ]{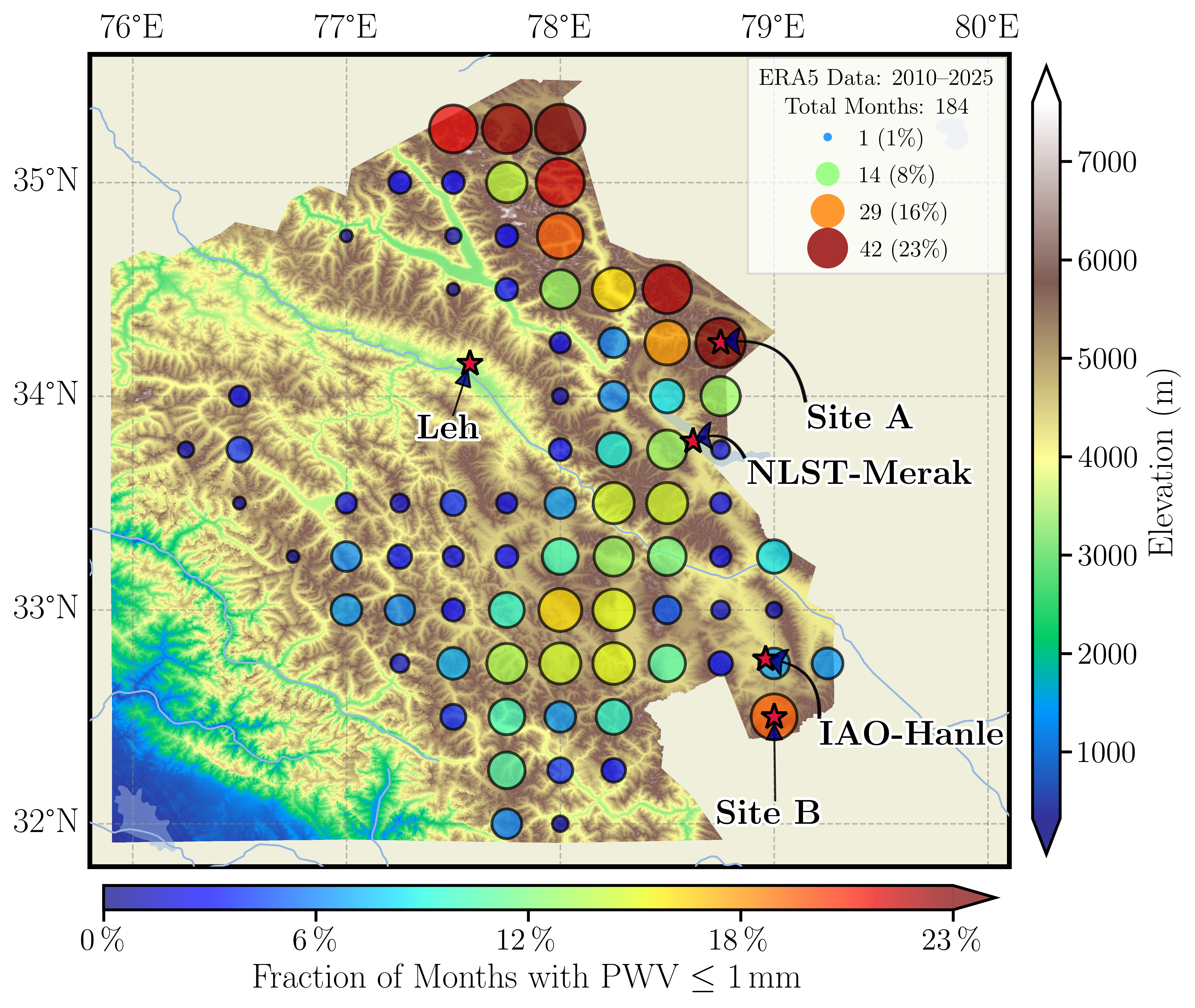}
\caption{Percent fraction of months in the Ladakh plateau from January 2010 to April 2025 with average monthly PWV of $1~\textrm{mm}$ or lower, displayed at native ERA5 resolution. The shaded relief map is the geographic area of interest. Large, dark red bubbles indicate highly promising areas with substantial fractions of months at or below the $1~\textrm{mm}$ threshold. Smaller, lighter blue bubbles indicate lower-performing areas. Regions without bubbles indicate no months meeting the $1~\textrm{mm}$ PWV criterion.}
\label{fig:bubbleplot}
\end{figure}

We now explore whether additional promising sites exist in the Ladakh plateau in addition to the established observatory locations at Hanle and Merak, where Hanle hosts the Indian Astronomical Observatory and Merak hosts the National Large Solar Telescope (NLST). We apply a PWV threshold of $1~\mathrm{mm}$ \citep{Shikhovtsev2025} to identify regions that are potentially suitable for hosting a submillimeter astronomical telescope.

The Ladakh plateau, characterized by its mean elevation of approximately $4500$ metres, encompasses varied topography including mountain peaks and valleys ranging from about $2800$ metres to over $6000$ metres in elevation. Due to this complex terrain, significant variations in PWV can exist within each pixel of the ERA5 grid. To address this inherent variability and identify promising regions systematically, we conduct a spatially coarse but comprehensive analysis at the native resolution of ERA5 across Ladakh and upper Himachal Pradesh, collectively referred to as the Ladakh plateau. Further pinpointing of optimal, specific local sites within these pixels would require detailed, on-site measurements, which are beyond the scope of this reanalysis-based investigation. Since there are no specifically identified sites with known latitude, longitude, and precise elevation beforehand, our approach relies on raw, monthly-averaged pressure level data directly provided by ERA5 at native spatial resolution. The dataset spans from January 2010 to April 2025 (184 months).

For this part of the study we use the native-grid method ($M3$) described in 
Section~\ref{subsec:pwv_calc}, employing the ERA5 surface pressure (\texttt{sp}) at 
each grid cell as the lower integration bound. It is important to emphasize that these 
results represent grid-cell average PWV profiles, which can smooth out sub-pixel variability; individual sites within a pixel may therefore have PWV values better or worse than the grid-cell mean.

For reference and clarity, the time-mean surface pressure values derived from the ERA5 monthly-averaged \texttt{sp} parameter at each pixel within our region of interest are shown in Fig. \ref{fig:pressure_map}. The $1\sigma$ range in $Pa$ shown inside each pixel provides insight into the typical variability of the surface pressure within that region. We observe that pixels located towards the North, Northeast, East, and Southeast boundaries of the study region consistently exhibit lower surface pressures compared to the central and western areas in the region of study. Consequently, we anticipate that these areas with lower surface pressures may correspond to lower PWV values, given that PWV is obtained from the integrated specific humidity content over the atmospheric column, which in turn depends significantly on surface pressure conditions.

Using ERA5 surface pressure values as the lower integration limit in Eq.~\ref{eq:PWV}, we compute monthly PWV values for each pixel over the study duration. Maintaining the benchmark PWV threshold of $1~\textrm{mm}$, we evaluate the frequency of months each pixel achieves PWV at or below this threshold from the full dataset of 184 months. Fig.~\ref{fig:bubbleplot} visualizes this metric, representing the fraction of months that satisfy this criterion across the Ladakh plateau. The color and size of the bubbles in this figure directly reflect each pixel’s duty cycle performance concerning this benchmark. The region of interest is depicted by relief map in the background \citep{Jarvis2008}. As expected, regions that statistically show the maximum number of months with PWV at or below $1~\textrm{mm}$ correlate well with those regions exhibiting low surface pressure in the pressure map of Fig. \ref{fig:pressure_map}.

Based on this analysis, we have several important remarks:
\begin{itemize}
\item Leh, which is the administrative capital of the Ladakh Union Territory, has zero months satisfying the $1~\textrm{mm}$ PWV threshold, making it unsuitable for a submillimeter telescope.
\item Hanle Pixel has 10 months and Merak Pixel has 14 months below the $1~\textrm{mm}$ threshold out of 184 total months. It may be noted that these are the corresponding pixels at native ERA5 resolution wherein the IAO-Hanle and NLST-Merak observatories lie.
\item Several pixels within the Ladakh plateau outperform Hanle and Merak Pixels in terms of the frequency of months showing PWV at or below $1~\textrm{mm}$. The best-performing pixels identified are (35.25$^{\circ}$N, 77.75$^{\circ}$E), (35.25$^{\circ}$N, 78.00$^{\circ}$E), and (34.50$^{\circ}$N, 78.50$^{\circ}$E), averaging approximately 42 months below the threshold. However, the logistics in terms of weather and access may make some of these locations impractical for further study. 

\item We denote two pixels around coordinates (34.25$^{\circ}$N, 78.75$^{\circ}$E) and (32.50$^{\circ}$N, 79.00$^{\circ}$E) as Site A and Site B respectively. Site A shows substantial promise, achieving $~\textrm{PWV}\leq1~\textrm{mm}$ in 43 out of 184 months, representative of the best site in Ladakh, and Site B achieving this benchmark in 35 months. Whereas Site A is challenging to access, Site B is likely more promising due to its proximity to IAO Hanle. 

\end{itemize}

\begin{figure}[htbp]
    \centering
    \includegraphics[width=\linewidth]{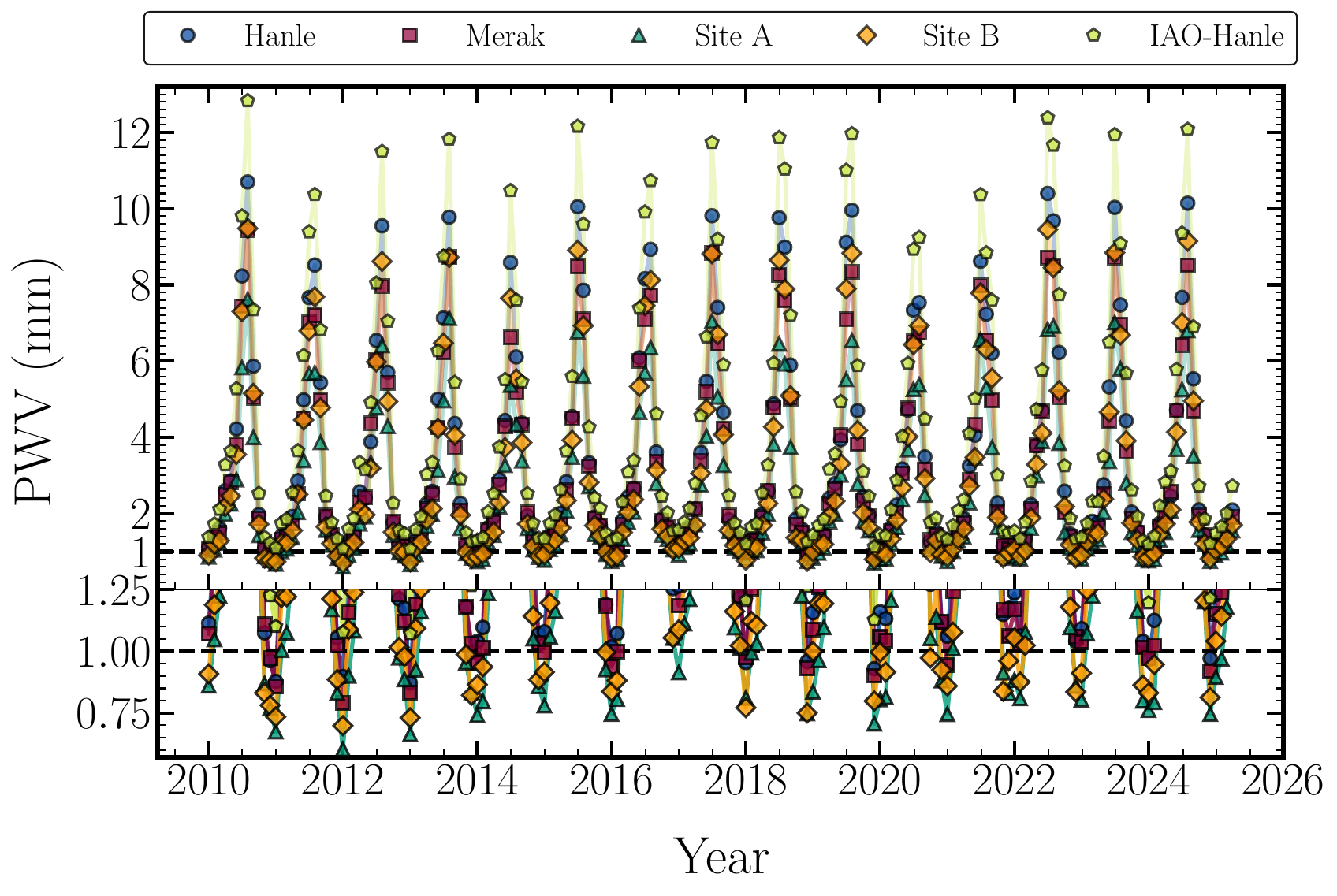}

    \vspace{-0.1em}
    \includegraphics[width=\linewidth]{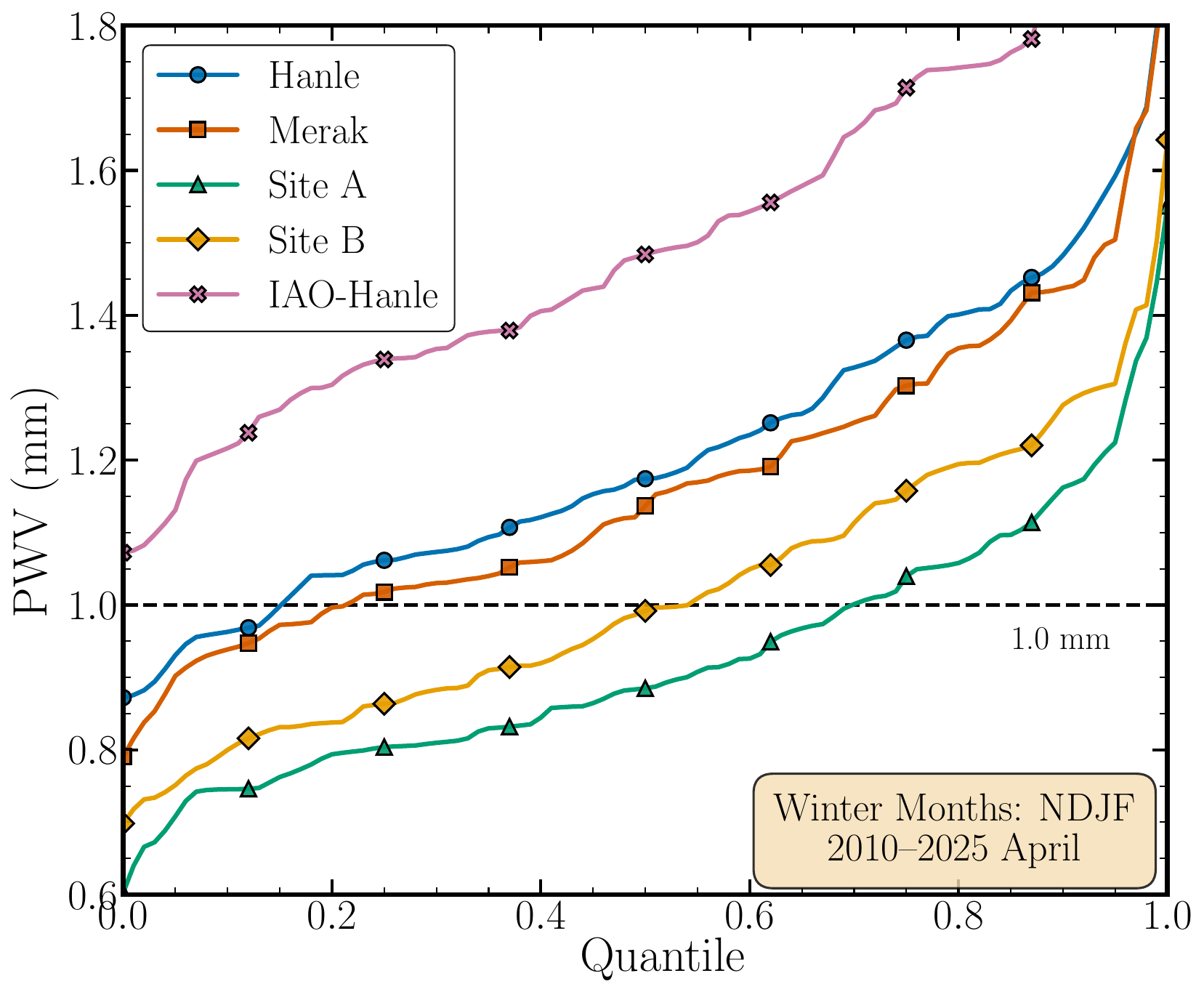}

    \caption{
    Top: Monthly precipitable water vapor (PWV) time series for the candidate sites on the Ladakh plateau over 2010--2025. The lower sub-panel shows a zoomed view around the PWV $\approx 1\,\mathrm{mm}$ benchmark commonly used for submillimeter observing conditions.
    Bottom: PWV quantile distributions for the same sites restricted to the winter months (NDJF) over 2010--2025.
    }
    \label{fig:combined_pwv_timeseries_quantiles}
\end{figure}

The 15-year monthly PWV distributions for Hanle Pixel, Merak Pixel, Site A, and Site B are illustrated in the upper panel of Fig.~\ref{fig:combined_pwv_timeseries_quantiles}. The lower panel zooms into the critical $1~\textrm{mm}$ threshold region. We note that Site A consistently shows the largest number of monthly occurrences at or below the threshold, followed by Site B. Hanle Pixel and Merak Pixel perform worse but show similar performance. Notably, the exact site of IAO-Hanle, which was evaluated separately with $M2$, has no month below the $1~\textrm{mm}$ threshold.

Based on expected climatic trends, favourable conditions predominantly occur during the winter months (November to February; NDJF). The lower panel of Fig.~\ref{fig:combined_pwv_timeseries_quantiles} presents the corresponding quantile distributions, illustrating the most favourable scenarios for each location. During these periods, Site~A maintains PWV \(\leq 1~\mathrm{mm}\) for approximately \(70\%\) of the time, Site~B for about \(50\%\), while the Hanle and Merak pixels show more modest performance, with occurrences of roughly \(15\%\) and \(20\%\), respectively. These duty-cycle fractions represent the proportion of NDJF months for which the monthly-mean PWV satisfies the 1\,mm threshold. While such conditions are often sustained over successive winter months, a clear understanding of the duration of uninterrupted observing hours would require time-resolved analysis at hourly or sub-daily resolution, which lies beyond the scope of this study. Nonetheless, this analysis provides sufficient motivation for evaluating the corresponding atmospheric transmittance and sky brightness temperatures, which are examined next to assess the overall suitability of these sites for submillimeter astronomical observations.

\subsection{Radiative Transfer Modeling}\label{sec:RT}

We convert the PWV values derived in Section~\ref{subsec:pwv_calc} into observing–band performance metrics using the \texttt{am} radiative–transfer code developed at the Smithsonian Astrophysical Observatory \citep{am_model}. The model solves the monochromatic radiative–transfer equation in a static, plane–parallel, layered atmosphere with line by line absorption and emission. Spectroscopic parameters are from the \texttt{HITRAN} database, which provides pressure- and temperature-dependent line strengths and broadening coefficients for the dominant microwave-to-submillimeter absorbers, including H$_2$O, O$_3$, and O$_2$ \citep{gordon_2022}. For each layer, pressure and temperature define the thermodynamic state, while species volume mixing ratios set extinction and emissivity; the line shape is treated with Voigt–Kielkopf broadening. The background field is the CMB ($T_0=2.7$\,K), and all results herein are reported at zenith (airmass\,=\,1) unless explicitly noted.

To couple ERA5 to \texttt{am}, we first construct layer-wise pressure and temperature profiles from the ERA5 \emph{mean monthly pressure-level fields} — that is, the monthly means at each pressure level for temperature, ozone, and specific humidity. For site-specific calculations (e.g., IAO–Hanle) we obtain a physically consistent surface pressure and vertical column at the known site elevation via the geopotential-height interpolation described in Section~\ref{subsec:pwv_calc}. For pixel-based analyses we retain the native ERA5 grid-cell surface pressure (\texttt{sp}). We then adopt the ERA5 monthly specific-humidity profile $q(p)$ directly, without applying any scaling; the integrated column thus equals the ERA5 PWV for each timestamp. This preserves the ERA5 vertical structure while ensuring column consistency between the meteorological input and the radiative–transfer calculation.

Unless stated otherwise, computations span $10$--$1000$\,GHz with a frequency step of $0.025$\,GHz. The computed outputs are the optical depth $\tau(\nu)$, the transmittance $t(\nu)\equiv e^{-\tau(\nu)}$, and the thermodynamic (blackbody-equivalent) brightness temperature $T_b(\nu)$ as functions of frequency. For inter-site comparisons we evaluate two representative atmospheric states per location: the median (50th percentile) over all months and the winter median restricted to November–February (NDJF), which capture typical and best-season conditions, respectively. For visual guidance in the submillimeter windows we also reference an 80\% transmittance line.

\begin{figure*}[htbp]
\centering
\includegraphics[trim={0.1cm 0.1cm 0.1cm 0.1cm},clip,width=\linewidth]{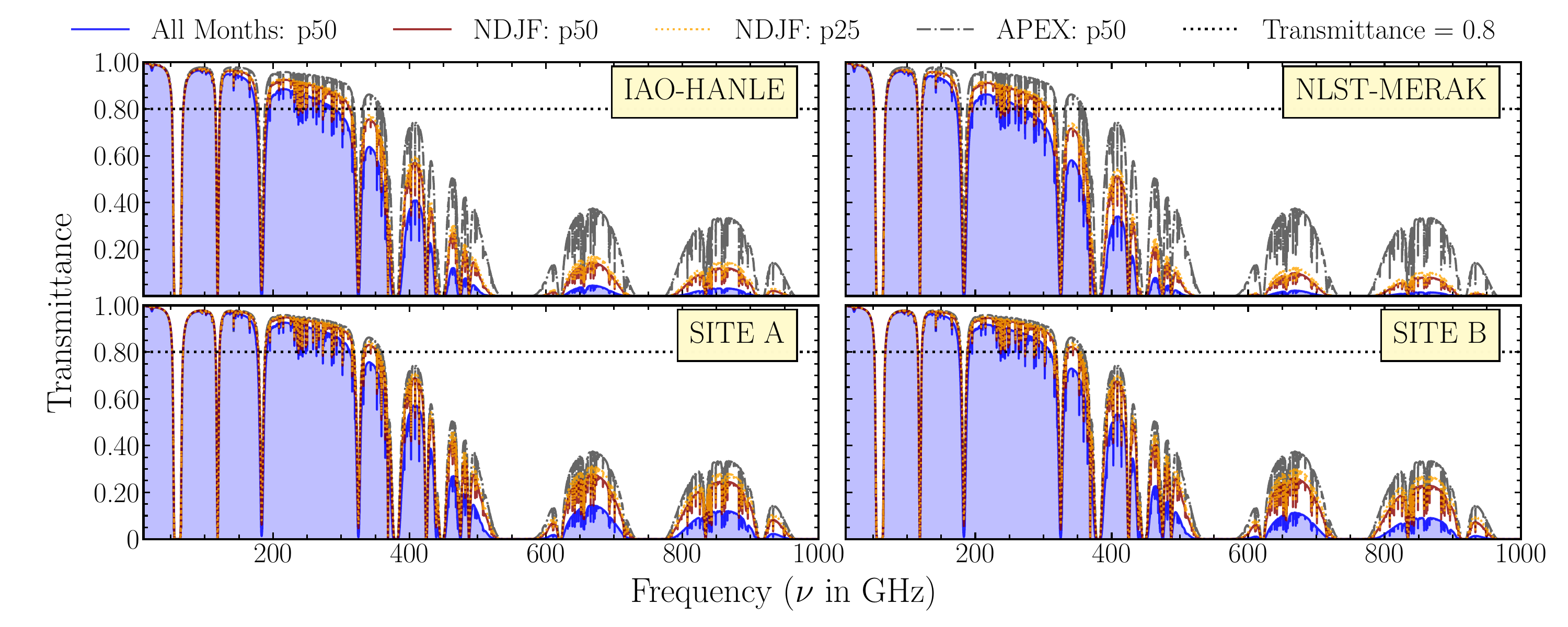}\\[2ex]
\includegraphics[trim={0.1cm 0.1cm 0.1cm 0.0cm},clip,width=\linewidth]{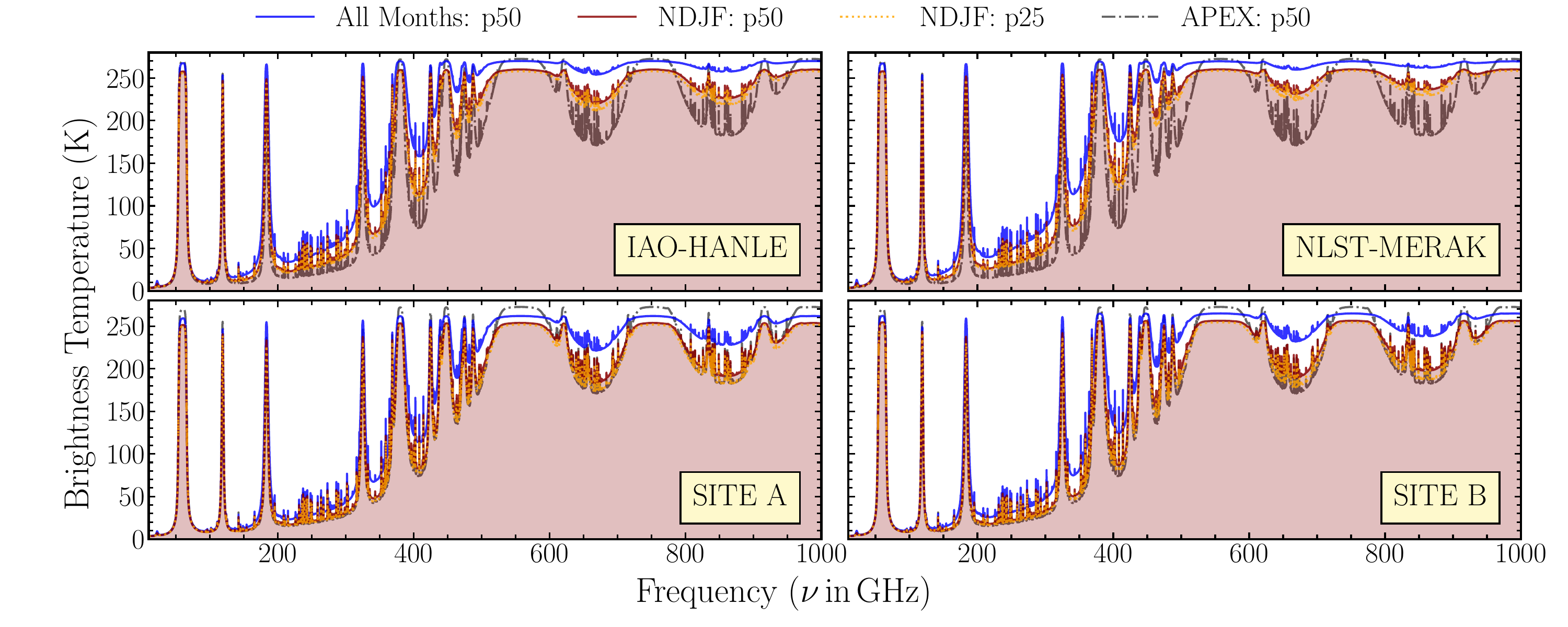}
\caption{
Top: Zenith transmittance spectra, $t_x(f)$, from \texttt{am} for the Hanle pixel, Merak pixel, Site~A, and Site~B. For each site, four curves are shown: the overall median (solid blue), NDJF median (solid red), NDJF 25th percentile (dotted orange), and the APEX median profile (dashed gray) from \citet{Valeria_2024}, based on MERRA-2 reanalysis. A horizontal line at $t_x=0.8$ marks a reference transmission level.
Bottom: Corresponding zenith sky brightness-temperature spectra, $T_b(f)$, for the same sites and atmospheric states. Lower $T_b$ values indicate reduced atmospheric emission and photon noise. Both panels follow identical color and line-style conventions for direct comparison.
}
\label{fig:transmittance_brightness_combined}
\end{figure*}

\begin{deluxetable}{lccc}
\tabletypesize{\small}
\tablewidth{0pt}   

\tablecaption{
Limiting frequency $\nu_{0.8}$ (GHz) corresponding to the highest frequency at
which the zenith atmospheric transmittance satisfies $t_\nu \ge 0.8$ for each site.
\label{tab:nu08_tx}
}

\tablehead{
\colhead{Site} &
\colhead{All p50} &
\colhead{NDJF p50} &
\colhead{NDJF p25}
}

\startdata
IAO--Hanle   & 290.18 & 312.77 & 314.43 \\
NLST--Merak  & 271.75 & 308.15 & 310.93 \\
Site~A       & 312.82 & 351.35 & 355.82 \\
Site~B       & 309.70 & 350.55 & 354.60 \\
APEX         & 358.02 & \nodata & \nodata \\
\enddata

\tablecomments{
``All p50'' denotes the median over all months; ``NDJF p50'' and ``NDJF p25''
denote the median and 25th percentile over NDJF (winter) months, respectively.
}
\end{deluxetable}

Across the four short-listed locations, winter (NDJF) conditions consistently yield higher transmittance and lower sky brightness than the full–year medians, reflecting the strong seasonal modulation of PWV. This behavior is clearly visible in the zenith spectra shown in Fig.~\ref{fig:transmittance_brightness_combined}, where the NDJF median and 25th percentile curves systematically outperform the annual median profiles. The APEX reference curve, derived from the median atmospheric profile for the APEX site (Cerro Chajnantor, Antofagasta, Chile) presented in \citet{Valeria_2024} using MERRA-2 reanalysis, provides a benchmark for a reference submillimeter site performance. A concise measure of high-frequency usability is provided by the limiting frequency $\nu_{0.8}$, which captures the highest frequency at which the zenith atmospheric transmittance $t_\nu$ remains above 0.8, excluding the narrow line centers of the 183 and 325\,GHz H$_2$O transitions and the 60\,GHz O$_2$ band and 119\,GHz isolated O$_2$ line. 

The corresponding values for each site and atmospheric state are summarised in Table~\ref{tab:nu08_tx}. Under typical conditions (All $p50$, i.e., median over all months of the year), Hanle and Merak reach $\nu_{0.8} \approx 290$ and 272\,GHz, respectively, while Site~A and Site~B extend to $\sim313$ and $\sim310$\,GHz. In winter medians (NDJF p50), this boundary expands substantially — Hanle and Merak to $\sim313$ and $\sim308$\,GHz, and Site~A and Site~B to $\sim351$\,GHz, indicating viability of the 325--345\,GHz window at the latter sites. Under especially dry winter states (NDJF p25), Site~A and Site~B approach $\nu_{0.8} \approx 356$ and 355\,GHz, whereas Hanle and Merak remain near $\sim314$ and $\sim311$\,GHz. Site~A exhibits the most favourable high-frequency windows, followed closely by Site~B, while Hanle and Merak degrade more rapidly above $\sim300$\,GHz as the 183 and 325\,GHz H$_2$O absorption features increasingly suppress the continuum transmission.

Looking at the upper panel of Fig.~\ref{fig:transmittance_brightness_combined} from an observational perspective leads to the same conclusion. Within the ALMA-defined frequency ranges, all sites show strong winter performance in Band~6 (211--275\,GHz), with transmittance values of $\sim0.89$--0.93, compared to $\sim0.95$ at APEX. At higher ALMA bands, Site~A and Site~B consistently perform better than Hanle and Merak, though they remain below APEX. In ALMA Band~7 (275--373\,GHz), Site~A and Site~B reach winter transmittance of $\sim0.76$ and $\sim0.75$, respectively, compared to $\sim0.80$ at APEX, while Hanle and Merak remain lower at $\sim0.69$ and $\sim0.65$. This separation becomes more pronounced in ALMA Band~8, where APEX shows transmission of $\sim0.40$, Site~A and Site~B reach $\sim0.34$ and $\sim0.33$, and Hanle and Merak drop to $\sim0.24$ and $\sim0.20$. In the 850\,$\mu$m window, Site~A and Site~B approach $\sim0.80$ and $\sim0.79$, compared to $\sim0.83$ at APEX, while Hanle and Merak remain at $\sim0.71$ and $\sim0.67$. In the more demanding 450 and 350\,$\mu$m windows, APEX retains transmission of $\sim0.25$--0.27, Site~A and Site~B maintain moderate values of $\sim0.16$--0.19, and Hanle and Merak fall below $\sim0.08$, indicating substantially reduced viability for sustained high-frequency submillimeter operation at those sites.

These trends also show up in the sky brightness temperature spectra (lower panel of Fig.~\ref{fig:transmittance_brightness_combined}), where the reduction in $T_b$ under winter conditions directly implies lower atmospheric emission and hence improved system sensitivity. Site~A consistently exhibits the largest performance margin across frequencies, followed closely by Site~B, whereas Hanle and Merak remain broadly comparable but increasingly constrained at higher frequencies.

\begin{figure}[htbp]
    \resizebox{\hsize}{!}{\includegraphics[scale=0.52]{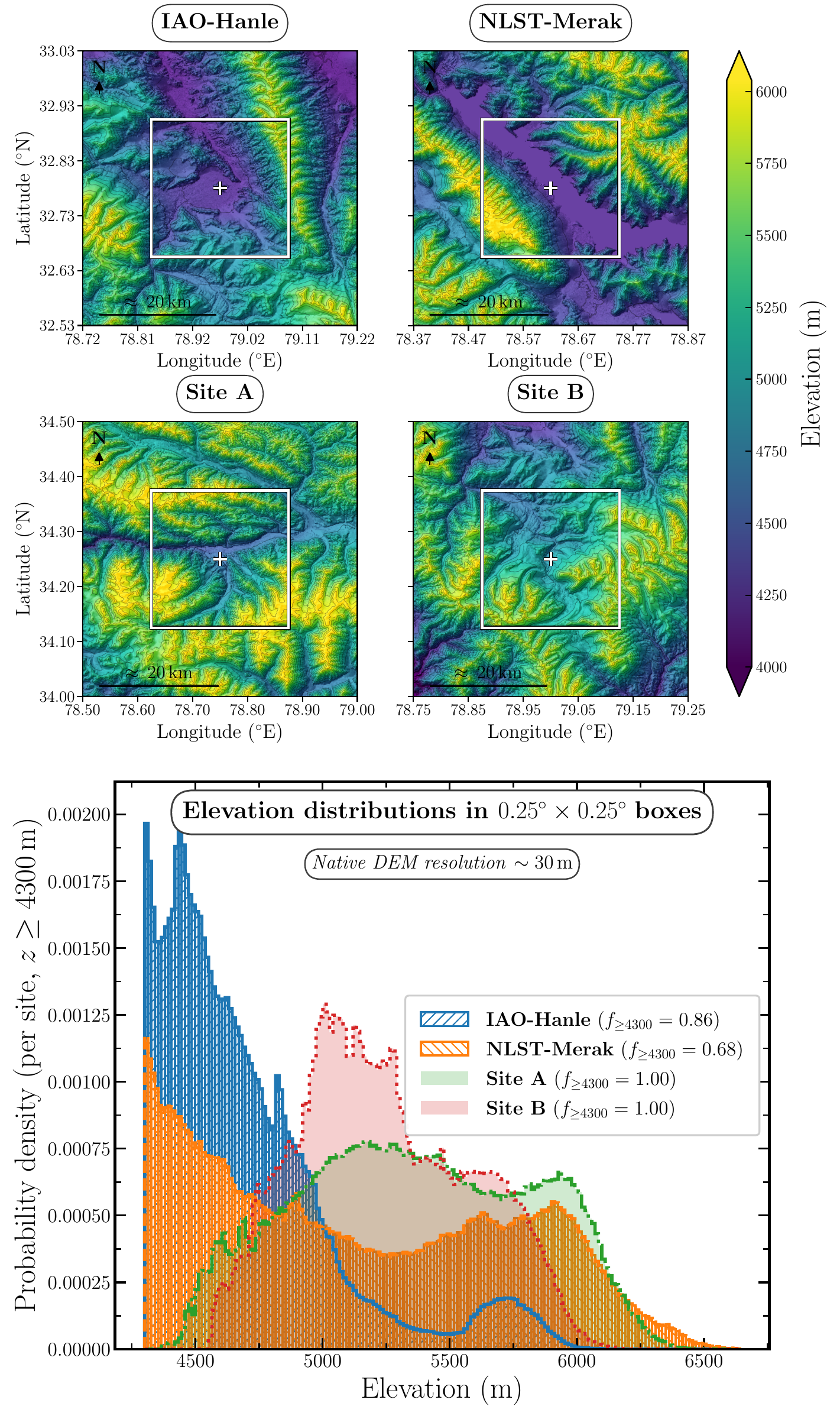}}
    \caption{
    Top: Digital elevation maps for the four candidate Ladakh site regions, each at $\sim$30\,m resolution shown in a $0.5^{\circ}\!\times\!0.5^{\circ}$ window ($\pm0.25^{\circ}$ about the site). A white cross marks the region center and the white square corresponds to the grid size in ERA5 ($0.25^{\circ}\!\times\!0.25^{\circ}$).
    Bottom: Probability density of surface elevation within $0.25^{\circ}\!\times\!0.25^{\circ}$ boxes centered on the four candidate sites. Each curve shows the normalized elevation histogram derived from the DEM (considering only grid cells above 4300\,m). $f_{\gtrsim 4300}$ denotes the fraction of DEM grid points within the region whose elevation exceeds 4300\,m.
    }
    \label{fig:combined_dem_hist}
\end{figure}

Section~\ref{subsec:hanle_validation} shows that ERA5 exhibits a positive PWV bias at IAO--Hanle relative to radiometer measurements. To remain conservative and internally consistent across all locations, including pixels lacking on-site data, no site-specific empirical corrections are applied to the transmittance calculations, unlike the approach adopted by \citet{Valeria_2024} through the use of an \texttt{nscale} factor. The reported transmittances can therefore be interpreted as lower bounds (and $T_b$ as upper bounds), given the sign of the ERA5 bias. The dominant sources of uncertainty include residual ERA5 moisture bias, representativeness over complex orography, and uncertainty in the vertical distribution of water vapor. By contrast, uncertainties in spectroscopic line strengths and pressure-broadening parameters are sub-dominant for our purposes, as indicated by recent high spectral-resolution heterodyne measurements at Chajnantor (see, e.g., \citealt{Pardo_2022,Pardo_2025}).

While a detailed forecasting and technology recommendation is beyond the scope of this paper, we provide an order-of-magnitude estimate with nominal values at 220 GHz. Assuming the formalism presented in \cite{carilli1999millimeter} we compute the effective system temperature referred to the top of the atmosphere $T^{eff}_{sys}$ as given by: 

\begin{equation}\label{eq:Tsys}
    T^{eff}_{sys} = e^{\tau}[T_{atm}\times(1-e^{-\tau})+T_{rx}]
\end{equation}

wherein $T_{atm}$ is the physical temperature of the atmosphere, $T_{rx}$ the noise temperature of the receiver, and $e^{-\tau}$ the transmittance. We adopt $T_{rx}$ to be 50 Kelvin for an SIS based receiver \citep{SIS_noisetemp} and $T_{atm}$ of 270 Kelvin over winter months in Ladakh, a conservative estimate based on \cite{ladakh_winter_temp}. When the transmittance is 0.9 we obtain $T^{eff}_{sys}$ to be $\sim 85$ Kelvin. This number degrades to $\sim 130$ Kelvin for a transmittance of 0.8. These are reasonable numbers and merit further investigation of site quality with dedicated instruments at candidate locations for observing band and technology selection for a science-class ground-based submillimeter telescope in Ladakh.

\section{Discussion and Conclusion}\label{section:discussion}

We have conducted a reanalysis-based study of the Ladakh plateau to examine site suitability for a future submillimeter telescope. The analysis focuses on precipitable water vapor (PWV) and asks whether locations in Ladakh have historically (in the ERA5 record) achieved PWV below a practical threshold of 1\,mm.

With the study conducted here, we make the following remarks.
\begin{itemize}
    \item The high-altitude peaks of Ladakh show promise for hosting a science-class submillimeter telescope and should be explored further.
    \item Using the grid–wise survey, we identify several regions in eastern Ladakh with a substantially higher fraction of time with PWV below 1 mm than the pixels containing Hanle and Merak.
    \item Two representative targets, Site~A $(34.25^\circ\mathrm{N},\,78.75^\circ\mathrm{E})$ and Site~B $(32.50^\circ\mathrm{N},\,79.00^\circ\mathrm{E})$, emerge as promising regions. Both outperform the Hanle and Merak pixels in the fraction of dry months over 2010–2025, more importantly, this advantage is strongest in winter (NDJF). Site~A performs best in our study in terms of PWV, and Site~B is noted to be the most promising for a detailed in situ study due to PWV-based performance and logistics and its proximity to Hanle.

\end{itemize}

Within each ERA5 pixel, terrain varies by several hundred metres, which implies meaningful variations in surface pressure and therefore in pixel-mean PWV. The top panel of Fig.~\ref{fig:combined_dem_hist} presents elevation maps within $0.5^\circ \times 0.5^\circ$ regions surrounding the four candidate sites, based on the $\sim$30 m–resolution DEM from \citet{ESA_CopernicusGDEM_2024}. Bottom panel of Fig.~ \ref{fig:combined_dem_hist} shows the corresponding per-site elevation distributions within $0.25^{\circ}\times0.25^{\circ}$ boxes (cells with elevations $\ge 4300$ m retained). The distributions are broad and multi-modal in places, and the legend fractions ($f_{\ge4300}$) emphasise how much of each box lies above high altitudes (IAO--Hanle $f_{\ge4300}\!=\!0.86$, NLST--Merak $0.68$, Site~A and Site~B $=1$). Elevation variations of 300–600 m within a single box imply several-percent differences in surface pressure, leading to non-negligible spread in column PWV. Consequently, finer-scale scouting within favorable pixels may identify local sites that are drier than the pixel-averaged conditions suggest \citep[see also for issues with complex orography across the Tibetan Plateau and surrounding mountain ranges:][]{Huang_2021,Ou2023NWTP}.

Our results are thus necessary but not sufficient evidence for site selection. Reanalysis characterizes climatological tendencies at coarse spatial resolution; it cannot substitute for targeted, on-site measurements that capture local topography, microclimate, and logistics. Practical constraints like access, infrastructure, regulations, etc. must be considered alongside atmospheric quality. Wind, seeing/turbulence, aerosol/sky-background and demographic constraints are also critical for final selection. As next steps, we propose: 

\begin{itemize}
    \item Re-examining the best-performing regions at higher time resolution and with additional reanalysis (e.g., ERA5-Land) for robustness.
    \item Conducting in situ PWV measurements at shortlisted sites over at least one full annual cycle to quantify and, if needed, correct ERA5’s positive moisture bias. Scalable options include GNSS-based PWV \citep{Bevis1992,Bevis1994} and compact microwave radiometers.
    \item Developing a strawman telescope configuration aligned with the science priorities and the atmospheric \emph{duty cycle} inferred here (e.g., band-averaged $tx$ and $T_b$ at 90, 150, 220, 270, 345\,GHz; zenith and representative elevations).
\end{itemize}
While these next steps will provide much needed detail, with the analysis presented in this work we conclude that the trans-Himalayan Ladakh region holds exciting potential as a promising candidate for a submillimeter observatory and merits further exploration.

\begin{acknowledgments}
T.S. gratefully acknowledges the Visiting Student Program (VSP) at the Raman Research Institute, Bengaluru, India, for providing the opportunity and support to carry out the research presented in this work.
\end{acknowledgments}

\bibliography{references}{}
\bibliographystyle{aasjournalv7}

\end{document}